\pdfoutput=1

\documentclass[english,journal]{IEEEtran}
\usepackage[T1]{fontenc}
\usepackage{ae}
\usepackage{aecompl}

\usepackage[latin9]{inputenc}
\usepackage{amsmath}
\usepackage{amssymb}
\usepackage{graphicx}

\makeatletter

\providecommand{\tabularnewline}{\\}

\makeatother

\usepackage{babel}
\begin{document}

\title{Multi-User Preemptive Scheduling For Critical Low Latency Communications
in 5G Networks}

\author{\IEEEauthorblockN{Ali A. Esswie$^{1,2}$,\textit{ Member, IEEE}, and\textit{ }Klaus
I. Pedersen\textit{$^{1,2}$, Senior Member, IEEE}\\
$^{1}$Nokia Bell-Labs, Aalborg, Denmark\\
$^{2}$Department of Electronic Systems, Aalborg University, Denmark
\\
Email: ali.esswie@nokia.com}}

\maketitle
$\pagenumbering{gobble}$
\begin{abstract}
5G new radio is envisioned to support three major service classes:
enhanced mobile broadband (eMBB), ultra-reliable low-latency communications
(URLLC), and massive machine type communications. Emerging URLLC services
require up to one millisecond of communication latency with 99.999\%
success probability. Though, there is a fundamental trade-off between
system spectral efficiency (SE) and achievable latency. This calls
for novel scheduling protocols which cross-optimize system performance
on user-centric; instead of network-centric basis. In this paper,
we develop a joint multi-user preemptive scheduling strategy to simultaneously
cross-optimize system SE and URLLC latency. At each scheduling opportunity,
available URLLC traffic is always given higher priority. When sporadic
URLLC traffic appears during a transmission time interval (TTI), proposed
scheduler seeks for fitting the URLLC-eMBB traffic in a multi-user
transmission. If the available spatial degrees of freedom are limited
within a TTI, the URLLC traffic instantly overwrites part of the ongoing
eMBB transmissions to satisfy the URLLC latency requirements, at the
expense of minimal eMBB throughput loss. Extensive dynamic system
level simulations show that proposed scheduler provides significant
performance gain in terms of eMBB SE and URLLC latency. 

\end{abstract}

\section{Introduction}

The standardization of the fifth generation (5G) new radio (NR) is
progressing with big momentum within the 3rd generation partnership
project (3GPP) community, to release the first 5G specifications {[}1-3{]}.
Ultra-reliable and low-latency communications (URLLC) is envisioned
as a key requirement of the 5G-type communications, to support broad
categories of many new applications from wireless industrial control,
autonomous driving, and to tactile internet {[}4{]}. URLLC services
require stringent latency and reliability levels, e.g., 1 ms at the
$1-10^{^{-5}}$reliability level {[}5{]}. Such a challenging latency
limit denotes that a URLLC packet which can not be transmitted and
successfully decoded before the URLLC latency deadline, is considered
as information-less and of no-use. 

Simultaneously achieving the requirements of extreme spectral efficiency
(SE) for enhanced mobile broadband (eMBB) services and ultra-low latency
for URLLC applications is a challenging problem {[}6{]}. Achieving
such URLLC latency demands more radio resources with ultra-low target
block error rate (BLER); though, it leads to a significant loss in
the network SE. Also, reserving dedicated resources for URLLC traffic
is spectrally inefficient due to its sporadic nature. 

To meet the stringent URLLC requirements, various studies have been
recently presented in the open literature. User-specific scheduling
with flexible transmission time intervals (TTIs) {[}7, 8{]} is recognized
as an enabler to achieve the URLLC latency limit, e.g., URLLC traffic
with a short TTI and eMBB with a longer TTI. However, the former increases
the aggregate overhead of the control channel. Additionally, different
configurations of microscopic and macroscopic diversity {[}9{]} are
proven beneficial for URLLC to significantly reduce the outage probability
of the signal-to-interference-noise-ratio (SINR). Advanced medium
access control enhancements {[}10{]} are also reported towards optimized
scheduling of URLLC traffic, including link adaptation filtering in
partly-loaded cells, dynamic and load-dependent BLER optimization.
Furthermore, preemptive scheduling {[}11, 12{]} is recently studied
to instantly schedule URLLC traffic within a shared channel, monopolized
by an ongoing eMBB transmission. Compared to existing studies, achieving
the URLLC latency requirements comes at the expense of a degraded
SE, e.g., high degrees of macroscopic diversity. Needless to say that
a flexible and multi-objective scheduling algorithm, which captures
the maximal system degrees of freedom (DoFs), is critical to reach
the best achievable URLLC-eMBB multiplexing gain. 

In this paper, a multi-user preemptive scheduling (MUPS) strategy
for densely populated 5G networks is proposed. MUPS aims to simultaneously
cross-optimize the network SE and URLLC latency. At each scheduling
TTI, MUPS scheduler assigns URLLC traffic a higher priority for immediate
scheduling without buffering. If sporadic URLLC traffic arrives at
the 5G general NodeB (gNB) during an arbitrary TTI, the gNB first
attempts to fit the URLLC packets within an ongoing eMBB transmission.
If the spatial DoFs are insufficient, the gNB decides to immediately
overwrite, i.e., preemptively schedule (PS), the physical resource
blocks (PRBs) over which URLLC users reported the best received SINR.
Compared to conventional PS scheduler, proposed MUPS utilizes the
spatial DoFs, offered by the transmit antenna array, to extract the
best achievable multiplexing gain, satisfying \textit{both}: URLLC
latency budget and eMBB throughput requirements.

Due to the complexity of the 5G NR system and the addressed problems,
performance evaluation is validated using advanced system level simulations
which offer high degree of realism and ensure reliable statistical
results. Those simulations are based on widely accepted models and
being calibrated with the 3GPP 5G NR assumptions {[}1-3{]}. 

This paper is organized as follows. Section II presents the system
model. Section III outlines the problem formulation and proposed MUPS
scheduler. Performance analysis appears in Section IV and the paper
is concluded in Section V.

\section{System Model}

We consider a downlink (DL) multi-user multiple-input multiple-output
(MU-MIMO) system, with $C$ cells. Each cell is equipped with $N_{t}$
transmit antennas while there are $K$-uniformly-distributed users
per cell, each with $M_{r}$ receive antennas. Users are dynamically
multiplexed through orthogonal frequency division multiple access
(OFDMA), and with 15 KHz sub-carrier spacing. There are two types
of DL traffic under evaluation: (1) URLLC time-sporadic traffic of
$Z$-bit finite payload per user with a Poisson point arrival process
$\lambda$, and (2) eMBB full buffer traffic with infinite payload.
The cell loading condition is described by $K_{URLLC}+K_{eMBB}=K$,
where $K_{URLLC}$ and $K_{eMBB}$ denote the average number of URLLC
and eMBB users per cell, respectively. URLLC traffic is scheduled
with a short TTI of 2 OFDM symbols (mini-slot of $0.143\,\textnormal{ms}$)
to meet the URLLC latency budget {[}1{]}. However, eMBB users are
scheduled with a long TTI of 14 OFDM symbols (slot of $1\,\textnormal{ms}$)
to maximize system SE.

A maximum MU subset $G\in K$, where $G_{c}\leq N_{t}$ is allowed
per PRB per cell, with equal power sharing. Thus, the received DL
signal at the $k^{th}$ user from the $c^{th}$ cell is given by

\[
\boldsymbol{\textnormal{y}}_{k,c}=\boldsymbol{\textnormal{\textbf{H}}}_{k,c}\boldsymbol{\textnormal{\textbf{V}}}_{k,c}s_{k,c}+\sum_{g\in G_{c},g\neq k}\mathit{\boldsymbol{\textnormal{\textbf{H}}}}_{k,c}\boldsymbol{\textnormal{\textbf{V}}}_{g,c}s_{g,c}
\]

\begin{equation}
+\sum_{j=1,j\neq c}^{C}\,\,\sum_{g\in G_{j}}\boldsymbol{\textnormal{\textbf{H}}}_{g,j}\boldsymbol{\textnormal{\textbf{V}}}_{g,j}s_{g,j}+\boldsymbol{\textnormal{\textbf{n}}}_{k},
\end{equation}
where $\boldsymbol{\textnormal{\textbf{H}}}_{k,c}\in\text{\ensuremath{\mathcal{C}}}^{M_{r}\times N_{t}},\forall k\in\{1,\ldots,K\},\forall c\in\{1,\ldots,C\}$
is the 3GPP spatial channel matrix seen by the $k^{th}$ user from
the $c^{th}$ cell, $\boldsymbol{\textnormal{\textbf{V}}}_{k,c}\in\mathcal{C}^{N_{t}\times1}$
and $s_{k,c}$ are the precoding vector (assuming a single stream
transmission) and the transmitted symbol, respectively. $\boldsymbol{\textnormal{\textbf{n}}}_{k}$
is the additive Gaussian white noise at the $k^{th}$ user. The first
summation in eq. (1) stands for the inter-user interference and the
second considers the inter-cell interference. The received signal
after applying the antenna combining vector $\boldsymbol{\textnormal{\textbf{U}}}_{k,c}\in\mathcal{C}^{M_{r}\times1}$
is given by 

\begin{equation}
\boldsymbol{\textnormal{y}}_{k,c}^{*}=\left(\boldsymbol{\textnormal{\textbf{U}}}_{k,c}\right)^{\textnormal{H}}\boldsymbol{\textnormal{y}}_{k,c},
\end{equation}
where $\left(.\right)^{\textnormal{H}}$ indicates the Hermitian transpose.
The antenna combining vector is designed based on the linear minimum
mean square error interference rejection combining (LMMSE-IRC) criteria
{[}13{]}, in order to project the received signal on a signal subspace
which minimizes the MSE, given by

\begin{equation}
\boldsymbol{\textnormal{\textbf{U}}}_{k,c}=\left(\boldsymbol{\textnormal{\textbf{H}}}_{k,c}\boldsymbol{\textnormal{\textbf{V}}}_{k,c}\left(\boldsymbol{\textnormal{\textbf{H}}}_{k,c}\boldsymbol{\textnormal{\textbf{V}}}_{k,c}\right)^{\textnormal{H}}+\boldsymbol{\textnormal{W}}\right)^{^{-1}}\boldsymbol{\textnormal{\textbf{H}}}_{k,c}\boldsymbol{\textnormal{\textbf{V}}}_{k,c},
\end{equation}
where $\boldsymbol{\textnormal{W}}=\text{\ensuremath{\mathbb{E}\left(\boldsymbol{\textnormal{\textbf{H}}}_{k,c}\boldsymbol{\textnormal{\textbf{V}}}_{k,c}\left(\boldsymbol{\textnormal{\textbf{H}}}_{k,c}\boldsymbol{\textnormal{\textbf{V}}}_{k,c}\right)^{\textnormal{H}}\right)}}+\sigma^{^{2}}\boldsymbol{\textnormal{I}}_{M_{r}}$
is the interference covariance matrix, $\text{\ensuremath{\mathbb{E}}}\left(.\right)$
denotes the statistical expectation, and $\boldsymbol{\textnormal{I}}_{M_{r}}$
is $M_{r}\times M_{r}$ identity matrix. The received SINR at the
$k^{th}$ user can be expressed as

{\small{}
\begin{equation}
\varUpsilon_{k,c}=\frac{p_{k}^{c}\left|\boldsymbol{\textnormal{\textbf{H}}}_{k,c}\boldsymbol{\textnormal{\textbf{V}}}_{k,c}\right|^{2}}{1+\underset{g\in G_{c},g\neq k}{\sum}p_{g}^{c}\left|\boldsymbol{\textnormal{\textbf{H}}}_{k,c}\boldsymbol{\textnormal{\textbf{V}}}_{g,c}\right|^{2}+\underset{j\in C,j\neq c\,}{\sum}\underset{g\in G_{j}}{\sum}p_{g}^{j}\left|\boldsymbol{\textnormal{\textbf{H}}}_{g,j}\boldsymbol{\textnormal{\textbf{V}}}_{g,j}\right|^{2}},
\end{equation}
}where $p_{k}^{c}$ is the transmission power of the $k^{th}$ user
in the $c^{th}$ cell. The per-user per-PRB data rate can then be
calculated as,

\begin{equation}
r_{k,rb}=\log_{2}\left(1+\frac{1}{\eta_{c}}\varUpsilon_{k,c}\right),
\end{equation}
where $\eta_{c}=\mathbf{card}(G_{c})$ is the MU rank on this PRB. 

Moreover, the link adaptation of the data transmission is based on
the frequency-selective channel quality indication (CQI) reports to
satisfy a target BLER. However, the CQI reports from the MU pairs
can be misleading since the calculation of the inter-user interference
and power sharing are not considered in the CQI estimation. Hence,
to stabilize the link adaptation process against MU variance, an offset
of $\delta$ dB is applied to the single-user (SU) CQI values before
the modulation and coding scheme (MCS) level is selected,

\begin{equation}
\varGamma_{\textnormal{MU}}=\varGamma_{\textnormal{SU}}-\delta,
\end{equation}
where $\varGamma_{\textnormal{MU}}$ and $\varGamma_{\textnormal{SU}}$
are the updated MU and reported CQI levels, respectively. Additionally,
due to the bursty nature of the URLLC traffic, it sporadically destabilizes
the reported CQI levels {[}10{]}, especially when an MU transmission
is not possible due to the fast varying interference patterns; otherwise,
the interference from the co-scheduled users contributes to stabilizing
the URLLC CQI levels. Thus, we further apply a sliding filter, e.g.,
a low pass filter, in order to smooth the instantaneous variation
rate of the CQI levels as follows, 

\begin{equation}
\partial(t)=\xi\varGamma_{\textnormal{MU}}+(1-\xi)\partial(t-1),
\end{equation}
where $\partial(t)$ is the MU CQI value to be considered for link
adaptation and MCS selection at the $t^{th}$ TTI, and $\xi\leq1$
is a tunable coefficient to specify how much weight should be given
to current reported CQI value. 

\section{Proposed Multi-User Preemptive Scheduling}

In this section, the concept of the proposed MUPS scheduler is introduced.
Under the 5G umbrella, there are multi user-specific, instead of network-specific,
objectives which need to be fulfilled simultaneously, e.g., eMBB SE
maximization, URLLC latency and BLER minimization as follows,

{\small{}
\begin{equation}
\text{\ensuremath{\forall}}k_{eMBB}\in\text{\ensuremath{\mathcal{K}}}_{eMBB}:\arg\underset{\text{\ensuremath{\mathcal{K}}}_{eMBB}}{\max}\sum_{k_{eMBB}=1}^{K_{eMBB}}\sum_{rb\in RB_{k}}r_{k,rb},
\end{equation}
}{\small \par}

{\small{}
\begin{equation}
\forall k_{URLLC}\in\text{\ensuremath{\mathcal{K}}}_{URLLC}:\arg\underset{\text{\ensuremath{\mathcal{K}}}_{URLLC}}{\min}\left(\beta\right),\,\beta\leq1\,\textnormal{ms},
\end{equation}
}{\small \par}

{\small{}
\begin{equation}
\forall k\in\text{\ensuremath{\mathcal{K}}}:\arg\underset{\mathcal{K}}{\min}(\psi),
\end{equation}
}where $\text{\ensuremath{\mathcal{K}}}_{eMBB}$ and $\text{\ensuremath{\mathcal{K}}}_{URLLC}$
denote the set of active eMBB and URLLC users, respectively. $\beta$
and $\psi$ indicate the URLLC latency at the $1-10^{^{-5}}$reliability
level and user BLER, respectively. This is a challenging and non-trivial
optimization problem, e.g., achieving Shannon SE requires infinite
latency budget. The proposed MUPS aims at achieving the maximum possible
system SE, while at the same time preserving the URLLC required latency. 

As shown in Fig. 1, if there is no incoming URLLC traffic at an arbitrary
TTI, MUPS assigns SU dedicated resources to incoming or buffered eMBB
traffic based on the proportional fair (PF) criteria as 

\begin{equation}
\Theta_{\textnormal{PF}}=\frac{r_{k,rb}}{\overline{r_{k,rb}}},
\end{equation}

\begin{equation}
k_{eMBB}^{*}=\arg\underset{\text{\ensuremath{\mathcal{K}}}_{eMBB}}{\max}\Theta_{\textnormal{PF}},
\end{equation}
where $\overline{r_{k,rb}}$ is the average delivered data rate of
the $k^{th}$ user. If incoming URLLC traffic is aligned at the start
of the current TTI, e.g., either it is a short URLLC or long eMBB
TTI, MUPS applies the weighted PF (WPF) criteria to instantly schedule
URLLC traffic with a higher priority on available resources as given
by

\begin{equation}
\Theta_{\textnormal{WPF}}=\frac{r_{k,rb}}{\overline{r_{k,rb}}}\alpha,
\end{equation}
where $\alpha$ is the scheduling coefficient and $\alpha_{URLLC}\gg\alpha_{eMBB}$.
Afterwards, MUPS schedules pending or new eMBB traffic on remaining
resources. 

If URLLC traffic arrives at the gNB during an eMBB TTI transmission
while scheduling resources are not available, gNB attempts to dynamically
multiplex the incoming short-TTI URLLC users within the ongoing long-TTI
eMBB transmissions, if there are sufficient spatial DoFs on this TTI.
The spatial DoFs represent the ability to jointly process several
signals between different sets of transmitters and receivers, if corresponding
channels are highly uncorrelated. Accordingly, URLLC users experience
no buffering overhead and then the URLLC latency budget can be satisfied.
If a successful pairing, i.e., MU URLLC-eMBB transmission over an
arbitrary PRB, is not possible, gNB will instantly overwrite the best
reported PRBs, known from the URLLC CQI reports, with the incoming
URLLC traffic. Thus, victim eMBB transmissions will exhibit a throughput
loss. 
\begin{figure}
\begin{centering}
\includegraphics[scale=0.5]{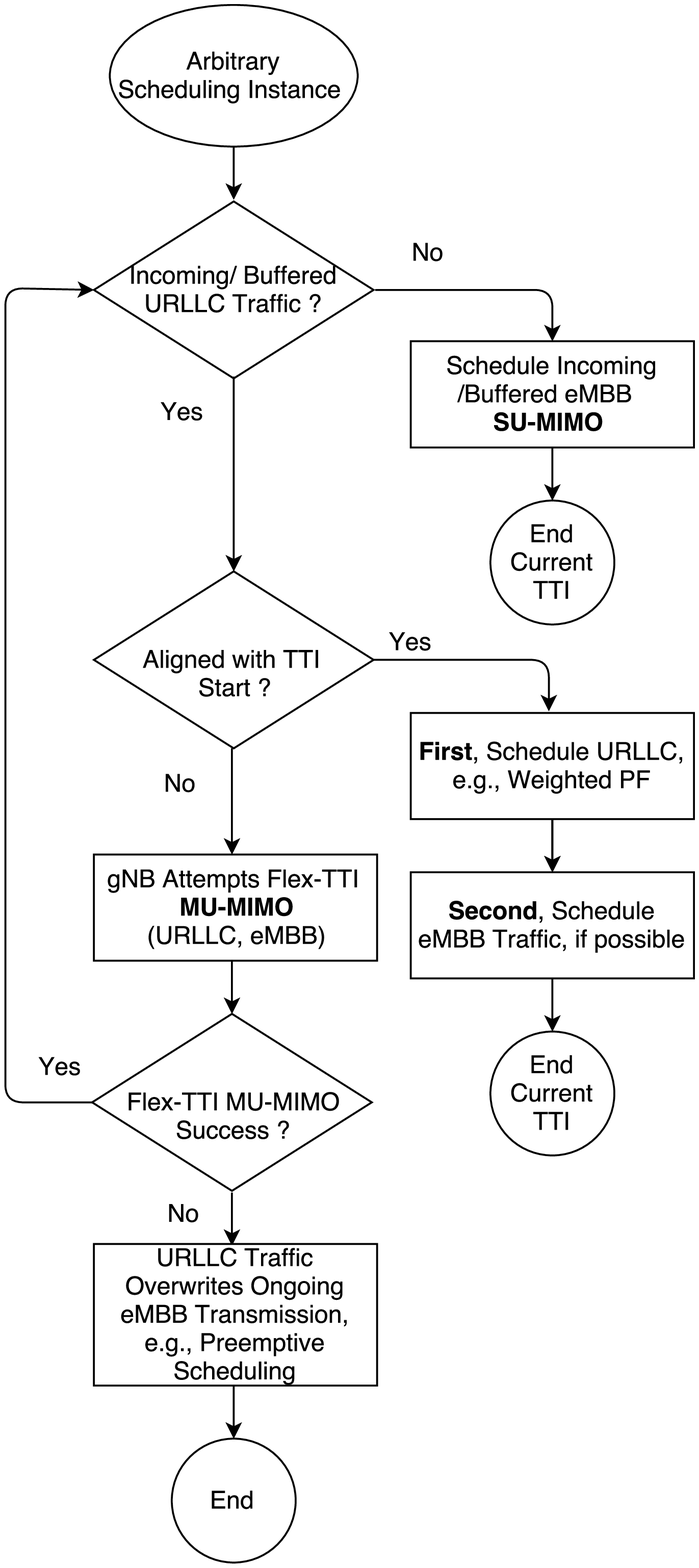}
\par\end{centering}
\centering{}{\small{}Fig. 1. Flow diagram of proposed MUPS scheduler.}{\small \par}
\end{figure}

For $N_{t}=8$ transmit antennas at the gNB, dual codebooks are defined
in LTE-Pro standards {[}14{]} for DL channel quantization at the user's
side, and are given by

\begin{equation}
\boldsymbol{\varLambda}_{1}=\left\{ \boldsymbol{v}{}_{1,1},\boldsymbol{v}{}_{1,2}\ldots,\boldsymbol{v}{}_{1,2^{B_{1}}}\right\} ,
\end{equation}

\begin{equation}
\boldsymbol{\varLambda}_{2}=\left\{ \boldsymbol{v}{}_{2,1},\boldsymbol{v}{}_{2,2}\ldots,\boldsymbol{v}{}_{2,2^{B_{2}}}\right\} ,
\end{equation}
where $\boldsymbol{v}{}_{i,j}$ denotes the $j^{th}$ codeword of
the $i^{th}$ codebook, $B_{1}$ and $B_{2}$ are the numbers of bits
of the two precoding matrix indices, reported from each user for the
gNB to select one codeword from each codebook. Each user projects
its estimated DL channel on both codebooks to select the closest possible
codewords as 

\begin{equation}
\widehat{\boldsymbol{v}}_{1}=\arg\underset{\boldsymbol{v}{}_{1}\in\boldsymbol{\varLambda}_{1}}{\max}\left\Vert \hat{\boldsymbol{\textnormal{\textbf{H}}}}\boldsymbol{\varLambda}_{1}\right\Vert ^{2},
\end{equation}

\begin{equation}
\widehat{\boldsymbol{v}}_{2}=\arg\underset{\boldsymbol{v}{}_{2}\in\boldsymbol{\varLambda}_{2}}{\max}\left\Vert \hat{\boldsymbol{\textnormal{\textbf{H}}}}\boldsymbol{\varLambda}_{2}\right\Vert ^{2},
\end{equation}
where $\left\Vert .\right\Vert $ denotes the 2-norm operation. The
final precoding vector at the gNB is obtained by the spatial multiplication
of both precoders, and is given by

\begin{equation}
\boldsymbol{\textnormal{\textbf{V}}}=\widehat{\boldsymbol{v}}_{1}\times\widehat{\boldsymbol{v}}_{2}.
\end{equation}

For a MU transmission on a given PRB, the zero-forcing (ZF) beamforming
is used to null the inter-user interference between the co-scheduled
pairs as expressed by

\begin{equation}
\boldsymbol{\textnormal{\textbf{V}}}_{\textnormal{MU}}=\left[\boldsymbol{\textnormal{\textbf{V}}}_{1}\ldots\boldsymbol{\textnormal{\textbf{V}}}_{\textnormal{G}}\right],
\end{equation}

\begin{equation}
\textbf{V}_{\textnormal{zf}}=\boldsymbol{\textnormal{\textbf{V}}}_{\textnormal{MU}}\left(\boldsymbol{\textnormal{\textbf{V}}}_{\textnormal{MU}}^{\textnormal{H}}\boldsymbol{\textnormal{\textbf{V}}}_{\textnormal{MU}}\right)^{-1}\textnormal{diag}\left(\sqrt{P}\right),
\end{equation}
where $\boldsymbol{\textnormal{\textbf{V}}}_{\textnormal{G}}$ and
$\textbf{V}_{\textnormal{zf}}$ present the precoder of the $g^{th}$
user enrolled in a MU-MIMO transmission and the ZF beamforming matrix,
where its column vectors are the data beamforming vectors of the MU
pairs. The MU transmission success is based on the maximization of
the Chordal distance between the ZF beamformers of the co-scheduled
users as follows,

\begin{equation}
\arg\underset{\boldsymbol{\textnormal{V}}_{\textnormal{eMBB}}\in\boldsymbol{\textnormal{\ensuremath{\mathcal{V}}}}_{\textnormal{eMBB}}}{\max}\,\,\textnormal{\textbf{d}}\left(\boldsymbol{\textnormal{V}}_{\textnormal{URLLC }},\boldsymbol{\textnormal{V}}_{\textnormal{eMBB}}\right),
\end{equation}
where $\boldsymbol{\textnormal{\ensuremath{\mathcal{V}}}}_{\textnormal{eMBB}}$
represents the set of ZF precoders of the eMBB active user set. The
Chordal distance is calculated as

\begin{equation}
\textnormal{\textbf{d}}\left(\boldsymbol{\textnormal{V}}_{\textnormal{URLLC}},\boldsymbol{\textnormal{V}}_{\textnormal{eMBB}}\right)=\frac{1}{\sqrt{2}}\left\Vert \boldsymbol{\textnormal{V}}_{\textnormal{URLLC}}\boldsymbol{\textnormal{V}}_{\textnormal{URLLC}}^{^{\textnormal{H}}}-\boldsymbol{\textnormal{V}}_{\textnormal{eMBB}}\boldsymbol{\textnormal{V}}_{\textnormal{eMBB}}^{^{\textnormal{H}}}\right\Vert .
\end{equation}

Upon MU pairing success, the aggregate achievable data rate on a given
PRB $r_{rb}$ is expressed by the sum rate of both co-scheduled URLLC
and eMBB users as

\begin{equation}
r_{rb}=\left(r_{eMBB}+r_{URLLC}-\Delta\right),
\end{equation}
where $\Delta$ represents the eMBB and URLLC SU rate loss due to
the MU inter-user interference. If a MU pairing is not possible, due
to either insufficient spatial DoFs or low number of active eMBB users,
the URLLC traffic immediately overwrites the PRBs over which it experiences
the best CQI levels. Thus, the eMBB users which have ongoing transmissions
on these PRBs suffer from throughput degradation. However, recovery
mechanisms can be arbitrarily considered not to include these PRBs
as part of the HARQ chase combining process and propagate errors,
e.g., consider these PRBs as information-less. Then, the sum rate
on victim PRBs can be expressed only by the achievable URLLC rate
as

\begin{equation}
r_{rb}=r_{URLLC}.
\end{equation}

For the sake of a fair URLLC latency evaluation, we compare the MUPS
performance with the preemptive-only scheduling (PS) {[}11{]}, where
incoming URLLC traffic always overwrites ongoing eMBB transmissions
without buffering, at the expense of the system SE. As it will be
discussed in Section IV, we demonstrate that a conservative multi-TTI
MU-MIMO transmission can be an attractive solution to approach both
URLLC latency and eMBB SE requirements. 

\section{Simulation Results}

Extensive dynamic system level simulations have been conducted, following
the 5G NR specifications in 3GPP {[}3{]}. The major simulation parameters
are listed in Table 1, where the baseline antenna setup is $8\times2$
unless otherwise mentioned. 

\begin{table}
\caption{Simulation Parameters.}
\centering{}%
\begin{tabular}{c|c}
\hline 
Parameter & Value\tabularnewline
\hline 
Environment & $\begin{array}{c}
\textnormal{3GPP-UMA,7 gNBs, 21 cells,}\\
\textnormal{500 meters inter-site distance}
\end{array}$\tabularnewline
\hline 
Channel bandwidth & 10 MHz, FDD\tabularnewline
\hline 
gNB antennas & 8, 16 and 64 Tx, 0.5$\lambda$\tabularnewline
\hline 
User antennas & 2, 8, 16 and 64 Rx, 0.5$\lambda$\tabularnewline
\hline 
User dropping & $\begin{array}{c}
\textnormal{uniformly distributed}\\
\textnormal{URLLC: 5 and 10 users/cell}\\
\textnormal{eMBB: 5 , 10 and 20 users/cell}
\end{array}$\tabularnewline
\hline 
User receiver & LMMSE-IRC\tabularnewline
\hline 
TTI configuration & $\begin{array}{c}
\textnormal{URLLC: 0.143 ms (2 OFDM symbols)}\\
\textnormal{eMBB: 1 ms (14 OFDM symbols)}
\end{array}$\tabularnewline
\hline 
MAC scheduler(s) & $\begin{array}{c}
\textnormal{URLLC: WPF, SU/MU-MIMO and PS}\\
\textnormal{eMBB: PF, and SU/MU-MIMO}
\end{array}$\tabularnewline
\hline 
CQI & periodicity: 5 ms, with 2 ms latency, $\xi=0.01$\tabularnewline
\hline 
HARQ & $\begin{array}{c}
\textnormal{asynchronous HARQ, chase combining}\\
\textnormal{HARQ round trip time = 4 TTIs}
\end{array}$\tabularnewline
\hline 
Link adaptation & $\begin{array}{c}
\textnormal{dynamic MCS}\\
\textnormal{target URLLC BLER : 1\%}\\
\textnormal{target eMBB BLER : 10\%}
\end{array}$\tabularnewline
\hline 
Traffic model & $\begin{array}{c}
\textnormal{\textnormal{URLLC: bursty, }Z=50 bytes, \ensuremath{\lambda=250} }\\
\textnormal{eMBB: full buffer}
\end{array}$\tabularnewline
\hline 
MU-MIMO setup & $\begin{array}{c}
\textnormal{MU beamforming : ZF}\\
\textnormal{MU rank \ensuremath{(\eta)} : 2}\\
\textnormal{CQI offset (\ensuremath{\delta}) : 3 dB}
\end{array}$\tabularnewline
\hline 
Link to system mapping & Mean mutual information per coded bit {[}11{]}\tabularnewline
\hline 
\end{tabular}
\end{table}

Fig. 2 shows the empirical complementary cumulative distribution function
(CCDF) of the URLLC latency statistics. We define the cell loading
state by $\varOmega=(K_{eMBB},K_{URLLC})$, where the aggregate URLLC
offered load per cell in bits/s is calculated as: $K_{URLLC}\times\lambda\times Z$.
Looking at the URLLC latency at the $10^{-5}$ level, both proposed
MUPS and PS schedulers achieve the 1-ms limit with $\varOmega=(5,5)$.
By increasing the system loading, e.g., $K_{eMBB}=10$ and $K_{URLLC}=10$,
the inter-cell interference becomes a dominant component and hence,
all schedulers suffer from throughput and latency degradation. Though,
MUPS scheduler still shows a decent URLLC latency performance, e.g.,
1.7 ms at $10^{-5}$ level. 

PF scheduler suffers from URLLC latency error floor since both URLLC
and eMBB users have the same scheduling priority, thus, URLLC large
queuing delays occur. WPF shows optimized URLLC latency; however,
it doesn't achieve the 1-ms limit since the sporadic URLLC traffic,
which is available during an eMBB TTI transmission, is buffered, i.e.,
not scheduled instantly, until the next available TTI opportunity.
\begin{figure}
\begin{centering}
\includegraphics[scale=0.59]{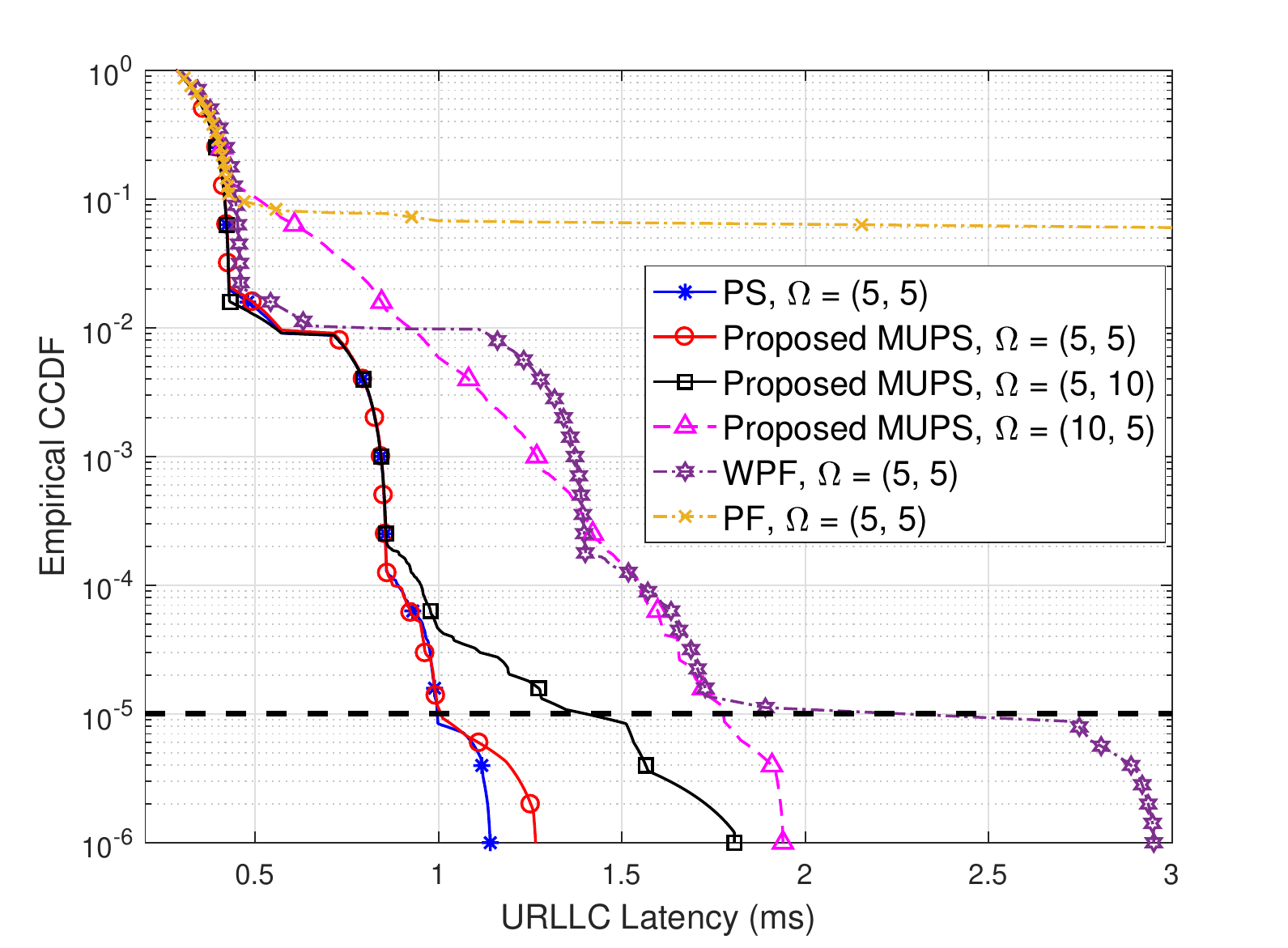}
\par\end{centering}
\centering{}{\small{}~~~~~Fig. 2. URLLC latency of MUPS, PS,
PF and WPF schedulers.}{\small \par}
\end{figure}

Fig. 3 shows the empirical CDF of the average cell throughput in Mbps
of the proposed MUPS and PS schedulers under different loading conditions.
Under all cell loading states, the MUPS scheduler shows significant
gain over PS scheduler, e.g., \textasciitilde{} $26.54\%$ gain with
$\varOmega=(20,5)$. MUPS scheduler exhibits a better system SE due
to: (1) the successful multi-TTI MU transmissions, and (2) reduction
in the number of the experienced PS scheduling events. For the same
number of the URLLC users $K_{URLLC}$, increasing the number of eMBB
users $K_{eMBB}$ significantly enhances the MU DoFs, hence, an incoming
URLLC user has higher probability to experience an immediate MU pairing
success, without falling back to SE-less-efficient PS scheduling.
Under such high $K_{eMBB}$ loading, MUPS scheduler attempts many
MU pairing success events; however, with limited MU gain due to the
aggregate level of inter-cell interference and the higher buffering
time. Thus, we also consider a modified version of the MUPS scheduler,
denoted as conservative MUPS (C-MUPS), where the URLLC-eMBB pairing
success becomes more restricted by the user spatial separation as

\begin{equation}
\left|\angle\left(\boldsymbol{\textnormal{V}}_{\textnormal{URLLC}}\right)-\angle\left(\boldsymbol{\textnormal{V}}_{\textnormal{eMBB}}\right)\right|^{o}\geq\theta,
\end{equation}
where $\theta$ is a predefined spatial separation threshold. Thus,
C-MUPS achieves lower number of MU attempts with further significant
MU gain, e.g., \textasciitilde{} $62$\% gain in average cell throughput
with $\varOmega=(20,5)$ and $\theta=60^{o}$, as shown in Fig. 3.

\begin{figure}
\begin{centering}
\includegraphics[scale=0.55]{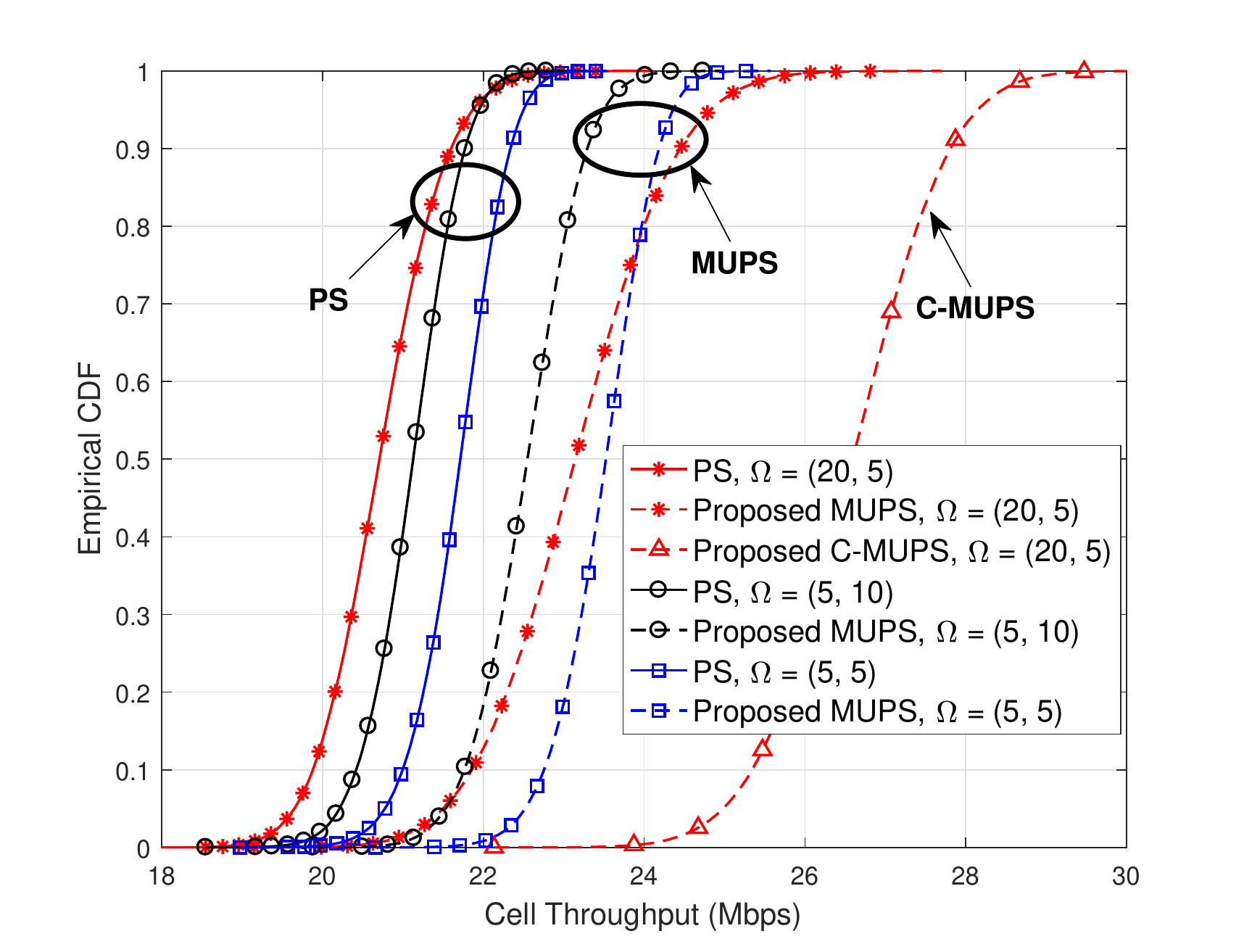}
\par\end{centering}
\centering{}{\small{}~~~~~~Fig. 3. Cell throughput of MUPS,
C-MUPS and PS schedulers.}{\small \par}
\end{figure}

As depicted in Fig. 4, it shows the average achievable MU throughput
increase with respect to average SU throughput. As can be noticed,
increasing $K_{URLLC}$ offers limited DoFs due to the short TTI length
of the URLLC users. Furthermore, increasing the URLLC load results
in more sporadic packet arrivals and hence, destabilizing the link
adaptation. Increasing the eMBB load offers great spatial DoFs per
each URLLC user. With C-MUPS, it shows that less MU success events
are experienced, e.g., 72\% instead of 95\% for MUPS with $\varOmega=(20,5)$;
however, further higher MU throughput is achieved. 
\begin{figure}
\begin{centering}
\includegraphics[scale=0.55]{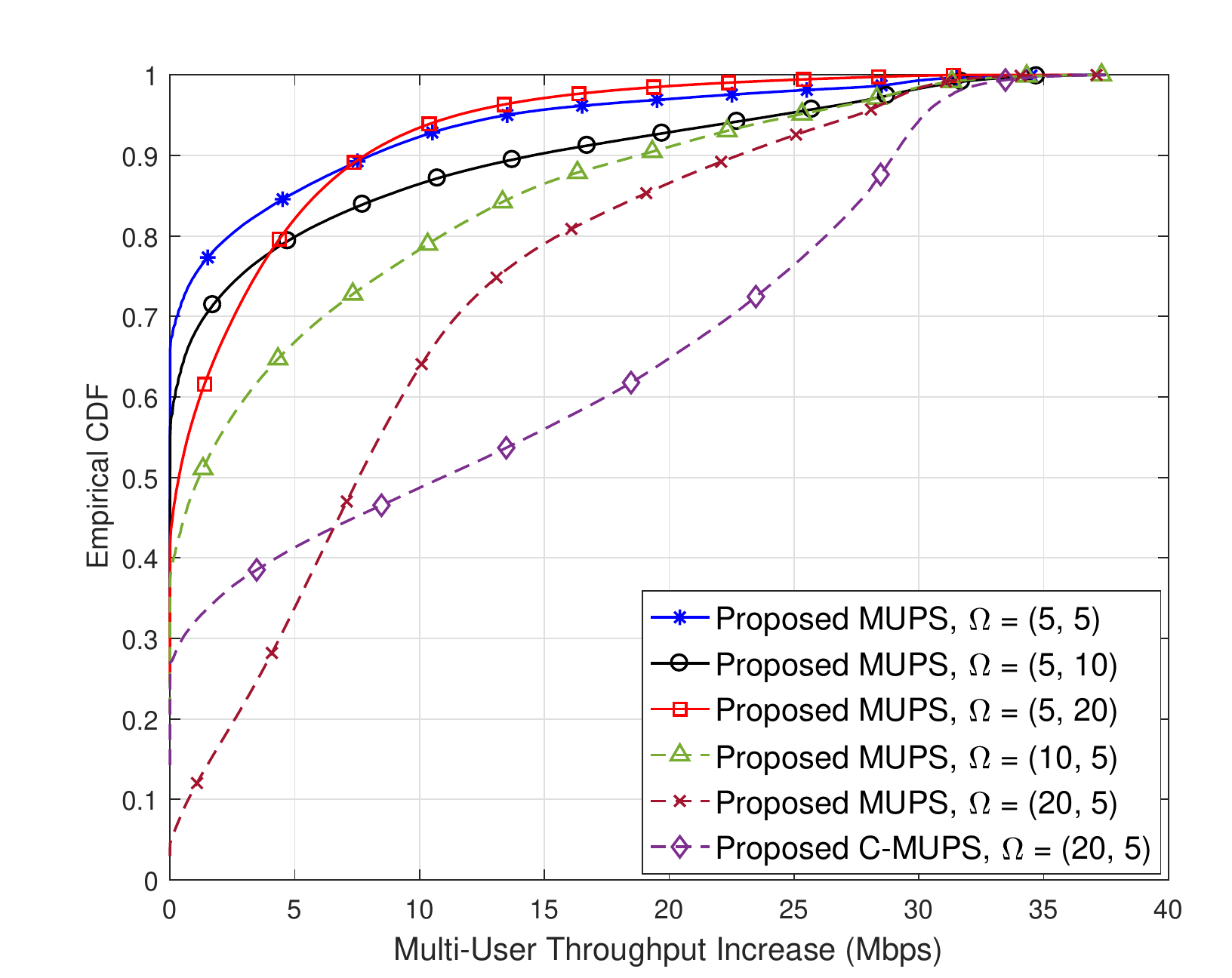}
\par\end{centering}
\centering{}{\small{}~~~~~~Fig. 4. MU throughput of the MUPS
and C-MUPS schedulers.}{\small \par}
\end{figure}

Examining the eMBB user performance, Fig. 5 presents a comparison
of the eMBB average user throughput. Proposed scheduler shows improved
eMBB user throughput, under all loading conditions. The gain in the
eMBB user throughput is strongly dependent on the levels of inter-cell
and inter-user interference. With light loading conditions, e.g.,
$\varOmega=(5,5)$, the MUPS scheduler experiences few successful
pairings with sub-optimal MU gain because of the insufficient available
spatial DoFs, e.g., due to the low value of $K_{eMBB}$ . On the opposite,
under heavy loading conditions, e.g., $\varOmega=(20,5)$, MUPS achieves
a higher number of successful MU pairings with higher MU gain as the
qr experiences few successful
pairings with sub-optimal MU gain because of the insufficient available
spatial DoFs, e.g., due to the low value of $K_{eMBB}$ . On the opposite,
under heavy loading conditions, e.g., $\varOmega=(20,5)$, MUPS achieves
a higher number of successful MU pairings with higher MU gain as the
quality of the MU transmission enhances with the number of active
eMBB users $K_{eMBB}$. 
\begin{figure}
\begin{centering}
\includegraphics[scale=0.55]{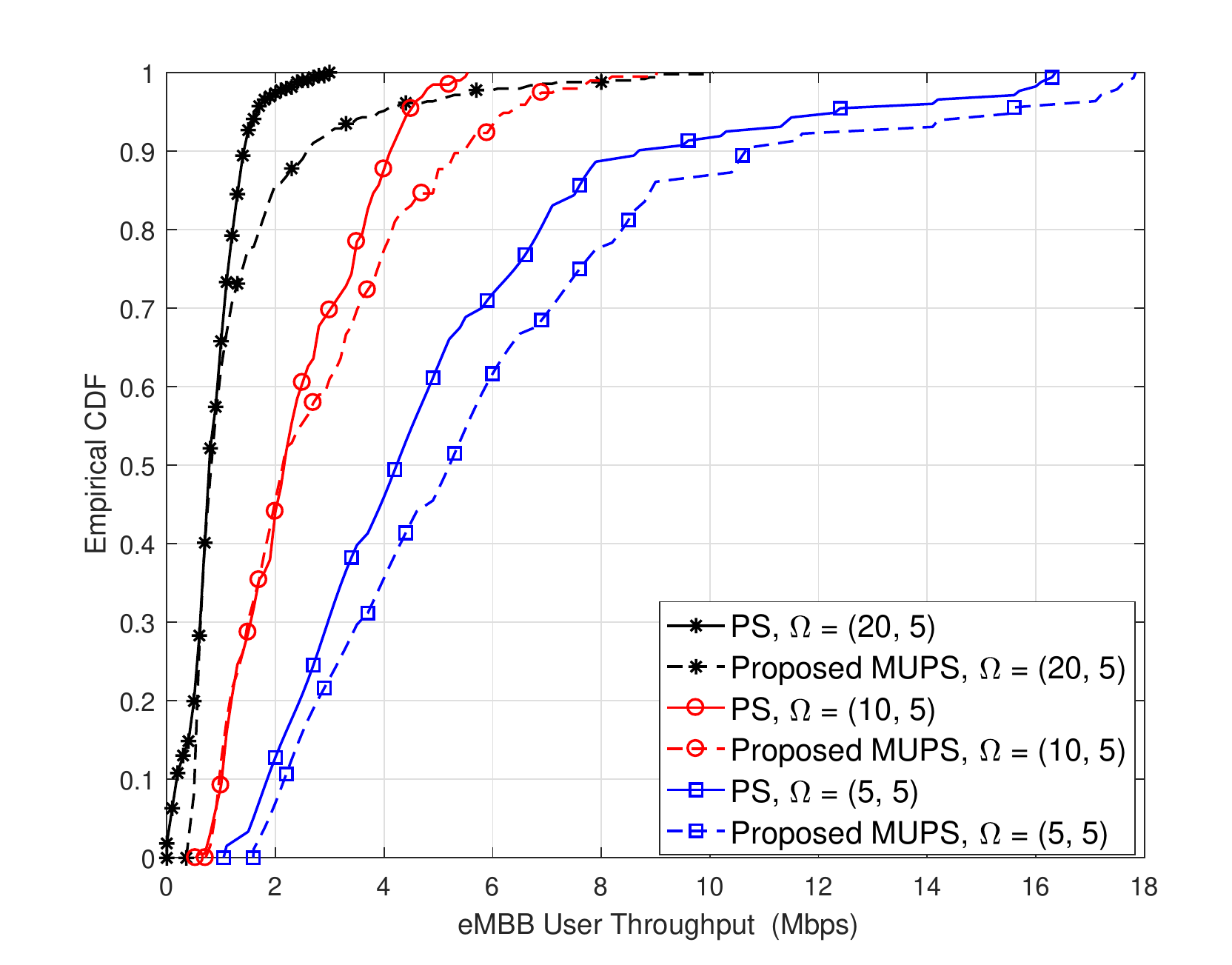}
\par\end{centering}
\centering{}{\small{}~~~~~~Fig. 5. eMBB user throughput of the
MUPS and PS schedulers.}{\small \par}
\end{figure}

Interestingly, the MU performance can be further improved with a larger
number of antennas, equipped at both transmitter and receiver. Channel
hardening {[}15, 16{]} denotes a fundamental channel phenomenon where
the variance of the channel mutual information shrinks as the number
of antennas grows, 

\begin{equation}
\sigma^{2}=\frac{1}{\min\left(N_{t},M_{r}\right)}\left(\frac{\left\Vert \boldsymbol{\textnormal{\textbf{H}}}\right\Vert ^{2}}{\mathbb{E}\left(\left\Vert \boldsymbol{\textnormal{\textbf{H}}}\right\Vert ^{2}\right)}\right).
\end{equation}

Consequently, the fading channel starts to act as a non-fading channel
where the channel eigenvalues become less sensitive to the actual
distribution of the channel entries. Thus, the channel hardens and
becomes much more directional on desired paths with less leakage on
the interfering paths, as shown in Fig. 6. As a result, both MU and
URLLC performance can be significantly improved. 
\begin{figure}
\begin{centering}
\includegraphics[scale=0.6]{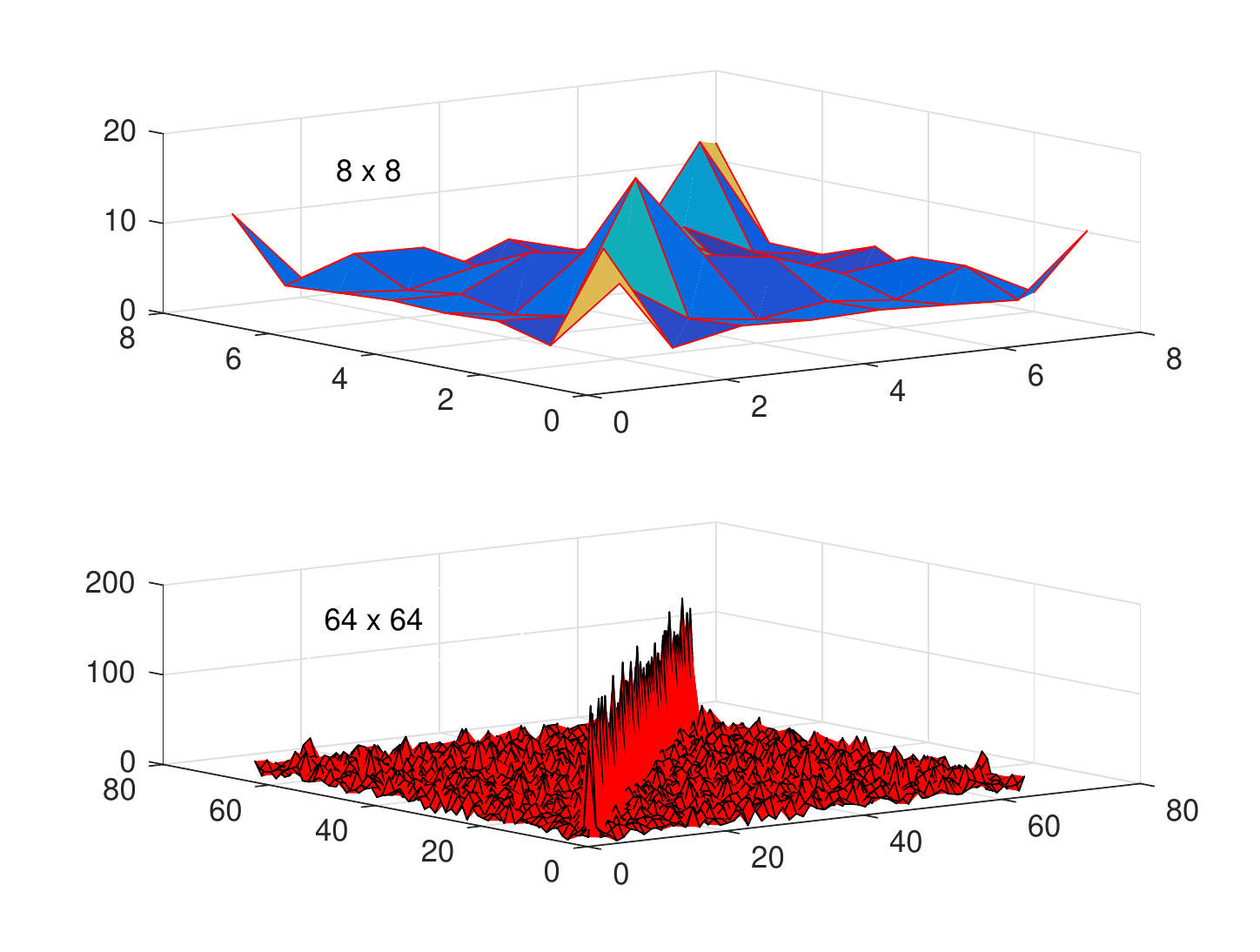}
\par\end{centering}
\centering{}{\small{}~~~~~~Fig. 6. Channel hardening of $\boldsymbol{\textnormal{H}}^{\textnormal{H}}\boldsymbol{\textnormal{H}}$
with $\left(N_{t},M_{r}\right)$ setup.}{\small \par}
\end{figure}

Fig. 7 introduces the received user SINR in dB, sampled over both
URLLC and eMBB users with $\varOmega=(20,5)$. For a fair performance
comparison, each user is assumed to feedback its serving cell with
the exact channel entries without quantization, since there is no
a standard quantization codebook for $N_{t}>8$ and $M_{r}>8$. The
channel is decomposed and fed-back by the singular value decomposition
{[}17{]} as: $\textbf{H}=\boldsymbol{\textnormal{\ensuremath{\mathcal{U}}}}\boldsymbol{\textnormal{\ensuremath{\varSigma}}}\boldsymbol{\textnormal{\ensuremath{\mathcal{V}}}}^{\textnormal{H}},$
where $\boldsymbol{\textnormal{\textbf{\ensuremath{\mathcal{U}}}}}\in\text{\ensuremath{\mathcal{C}}}^{M_{r}\times M_{r}}$
and $\boldsymbol{\text{\textbf{\ensuremath{\mathcal{V}}}}}\in\text{\ensuremath{\mathcal{C}}}^{N_{t}\times N_{t}}$
are unitary matrices and $\boldsymbol{\varSigma}\in\text{\ensuremath{\mathfrak{N}}}^{M_{r}\times N_{t}}$
is the channel singular matrix. The received user SINR levels are
significantly enhanced with the number of antennas due to the channel
hardening effect. Consequently, further more MU successful pairing
events can be achieved with sufficient spatial separation. 
\begin{figure}
\begin{centering}
\includegraphics[scale=0.6]{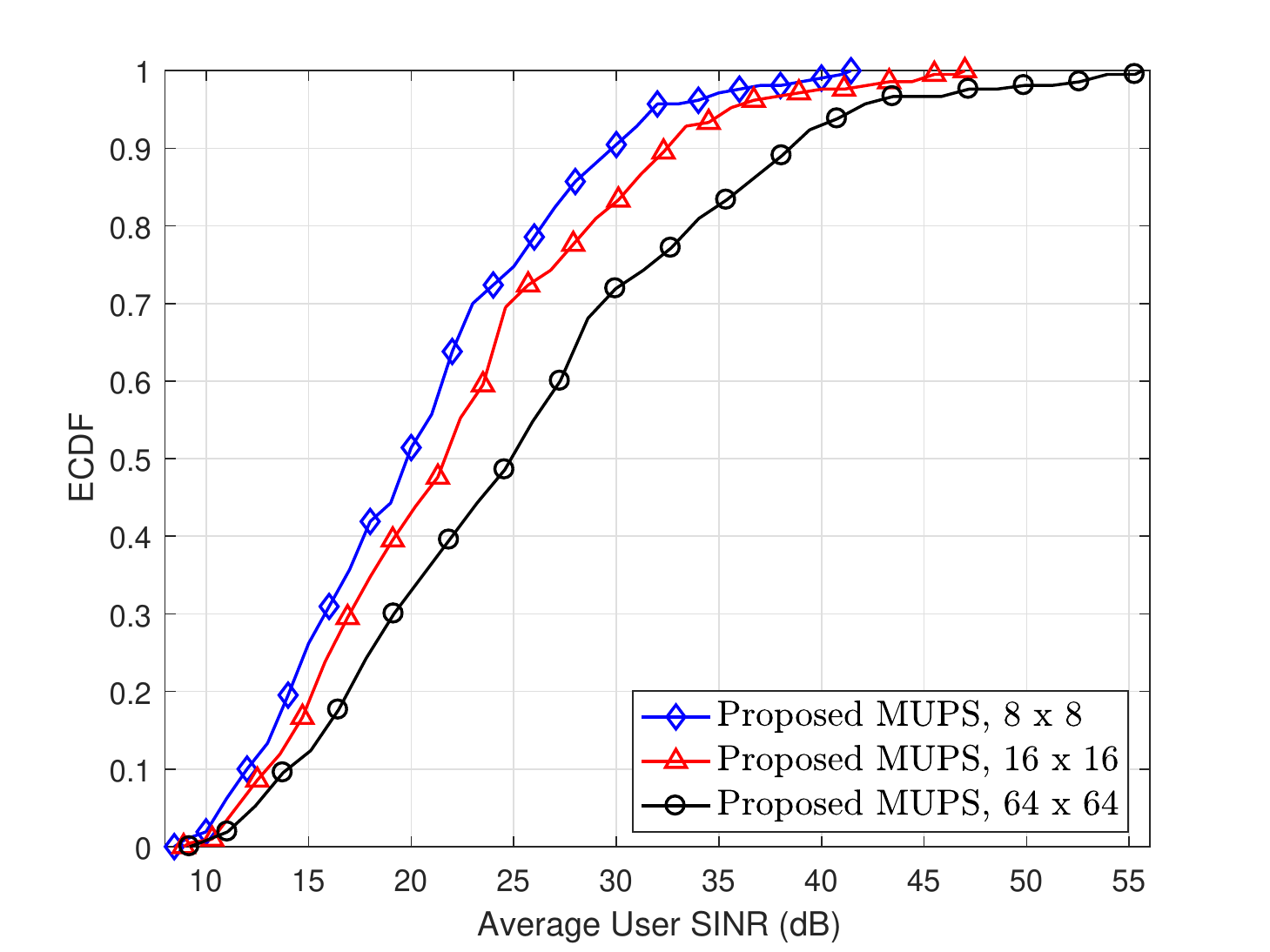}
\par\end{centering}
\centering{}{\small{}~~~~~~Fig. 7. User received SINR with $\left(N_{t},M_{r}\right)$
setup.}{\small \par}
\end{figure}

\section{Conclusion}

In this work, a joint multi-user preemptive scheduler (MUPS) has been
proposed for densely populated 5G networks. Proposed scheduler operates
efficiently with different traffic types, e.g., full buffer enhanced
mobile broadband (eMBB) and sporadic ultra-reliable low-latency communication
(URLLC) traffic. MUPS cross-optimizes the network performance such
that the maximum possible spectral efficiency and ultra low latency
are simultaneously achievable. Using extensive system level simulations,
the proposed scheduler provides significant performance gain, e.g.,
\textasciitilde{} $62$\% gain in average cell throughput, under different
network configurations. The performance of the MUPS scheduler is shown
to improve with the number of eMBB users until the interference levels
become dominant. Hence, proposed conservative MUPS shows further enhanced
MU gain by limiting the inter-user interference. Furthermore, increasing
the number of antennas is shown to harden the wireless channel and
thus, further improved URLLC performance can be satisfied. A detailed
study on the robustness of the URLLC performance under such a scenario
will be considered in a future work.


\begin{thebibliography}{10}
\bibitem[1]{key-1}NR and NG-RAN overall description; Stage-2 (Release
15), 3GPP, TS 38.300, V2.0.0, Dec. 2017. 

\bibitem[2]{key-2}Study on new radio access technology; Radio access
architecture and interfaces (Release 14), 3GPP, TR 38.801, V14.0.0,
March 2017.

\bibitem[3]{key-3}Study on scenarios and requirements for next generation
access technologies (Release 14), 3GPP, TR 38.913, V14.3.0, June 2016.

\bibitem[4]{key-4}IMT vision \textendash{} \textquotedblleft Framework
and overall objectives of the future development of IMT for 2020 and
beyond\textquotedblright , international telecommunication union (ITU),
ITU-R M.2083-0, Feb. 2015.

\bibitem[5]{key-5}E. Dahlman et al., \textquotedbl{}5G wireless access:
requirements and realization,\textquotedbl{}\textit{ IEEE Commun.
Mag.}, vol. 52, no. 12, pp. 42-47, Dec. 2014.

\bibitem[6]{key-6}B. Soret, P. Mogensen, K. I. Pedersen and M. C.
Aguayo-Torres, \textquotedbl{}Fundamental tradeoffs among reliability,
latency and throughput in cellular networks,\textquotedbl{} \textit{in
Proc. IEEE Globecom}, Austin, TX, 2014, pp. 1391-1396.

\bibitem[7]{key-7}K. I. Pedersen, G. Berardinelli, F. Frederiksen,
P. Mogensen and A. Szufarska, \textquotedbl{}A flexible 5G frame structure
design for FDD cases,\textquotedbl{} \textit{IEEE Commun. Mag.}, vol.
54, no. 3, pp. 53-59, March 2016.

\bibitem[8]{key-8}Q. Liao, P. Baracca, D. Lopez-Perez and L. G. Giordano,
\textquotedbl{}Resource scheduling for mixed traffic types with scalable
TTI in dynamic TDD systems,\textquotedbl{} \textit{in Proc. IEEE Globecom},
Washington, DC, 2016, pp. 1-7.

\bibitem[9]{key-9}G. Pocovi, B. Soret, M. Lauridsen, K. I. Pedersen
and P. Mogensen, \textquotedbl{}Signal quality outage analysis for
URLLC in cellular networks,\textquotedbl{} \textit{in Proc. IEEE Globecom},
San Diego, CA, 2015, pp. 1-6.

\bibitem[10]{key-10}G. Pocovi, B. Soret, K. I. Pedersen and P. Mogensen,
\textquotedbl{}MAC layer enhancements for ultra-reliable low-latency
communications in cellular networks,\textquotedbl{} \textit{in Proc.
IEEE ICC}, Paris, 2017, pp. 1005-1010.

\bibitem[11]{key-11}K.I. Pedersen, G. Pocovi, J. Steiner, and S.
Khosravirad, \textquotedblleft Punctured scheduling for critical low
latency data on a shared channel with mobile broadband,\textquotedblright{}\textit{
in Proc. IEEE VTC}, Toronto, 2017, pp. 1-6.

\bibitem[12]{key-12}G. C. Buttazzo, M. Bertogna and G. Yao, \textquotedbl{}Limited
preemptive scheduling for real-time systems: a survey,\textquotedbl{}
\textit{IEEE Trans. Ind. Informat.}, vol. 9, no. 1, pp. 3-15, Feb.
2013.

\bibitem[13]{key-13}Y. Ohwatari, N. Miki, Y. Sagae and Y. Okumura,
\textquotedbl{}Investigation on interference rejection combining receiver
for space\textendash frequency block code transmit diversity in LTE-advanced
downlink,\textquotedbl{} \textit{IEEE Trans. Veh. Technol.}, vol.
63, no. 1, pp. 191-203, Jan. 2014.

\bibitem[14]{key-14}Evolved universal terrestrial radio access (E-UTRA);
Physical layer procedures (Release 12), 3GPP, TS 36.213, V12.4.0,
Feb. 2015

\bibitem[15]{key-15}T. L. Narasimhan and A. Chockalingam, \textquotedblright Channel
hardening-exploiting message passing receiver in large-scale MIMO
systems,\textquotedblright{} \textit{IEEE J. Sel. Topics Signal Process.},
vol. 8, no. 5, pp. 847-860, Oct. 2014.

\bibitem[16]{key-16}A. A. Esswie, M. El-Absi, O. A. Dobre, S. Ikki
and T. Kaiser, \textquotedbl{}A novel FDD massive MIMO system based
on downlink spatial channel estimation without CSIT,\textquotedbl{}
\textit{in Proc. IEEE ICC}, Paris, 2017, pp. 1-6.

\bibitem[17]{key-17}D. W. Browne, M. W. Browne and M. P. Fitz, \textquotedbl{}CTH07-4:
singular value decomposition of correlated MIMO channels,\textquotedbl{}
\textit{in Proc. IEEE Globecom}, San Francisco, CA, 2006, pp. 1-6.
\end{thebibliography}
\end{document}